\documentstyle[12pt]{article}
\raggedright   %
\newfont{\tnbf}{cmbx10}
\newfont{\ninebf}{cmbx9}
\def\pr{\prime}

\def\and{\th\th\&\th\th}

\def\a{\alpha}
\def\b{\beta}

\def\Hilbert#1{{\cal H}_{#1}}
\def\ket#1{|\,#1\,\rangle}
\def\bra#1{\langle\, #1\,|}
\def\braket#1#2{\langle\, #1\,|\,#2\,\rangle}
\def\matrix#1#2#3{\langle\,#1\, |\, #2\, |\,#3\,\rangle}

\def\expect#1{\langle\, #1\, \rangle}

\def\th{\thinspace}

\def\thirty{\hbox to \hsize{\hfill\rule[5pt]{2.5cm}{0.5pt}\hfill}}

\def\half{\frac{1}{2}}

\def\squarthtwo{\frac{1}{\sqrt 2}}

\def\Id{\bf 1}

\def\dee#1#2{\frac{\partial #1}{\partial #2}}

\begin{document}
\null

\pagestyle{myheadings}
\markright{\ninebf Begging the Signalling Question\th\hrulefill\th\th}
\thispagestyle{plain}

\noindent{\large\bf Begging the Signalling Question:\\
\medskip
\noindent Quantum Signalling and the Dynamics of Multiparticle 
Systems.}\footnote{\small Submitted to the Proceedings of the Meeting of 
the Society for Exact Philosophy, 1999.}

\vskip36pt
\indent\indent\parbox{3in}{{\bf Kent A. Peacock}\\[6pt]

  {\sl Department of Philosophy,\\
  University of Lethbridge,       \\
  4401 University Drive,   \\
  Lethbridge, Alberta, Canada.   T1K 3M4\\
  kent.peacock@uleth.ca}}

\bigskip\bigskip
\indent\indent\parbox{3in}{{\bf Brian Hepburn}\\[6pt]

  {\sl Departments of Philosophy and Physics,\\
  University of Lethbridge, \\
  4401 University Drive,      \\
  Lethbridge, Alberta, Canada.  T1K 3M4 \\
  hepbbs@uleth.ca}}

\bigskip\medskip\bigskip
{\baselineskip=14pt
\noindent{\bf  ABSTRACT} \\[6pt]
  \noindent{The abundant experimental
  confirmation of Bell's Theorem has made a
  compelling case for the nonlocality of quantum mechanics (QM), in the
  precise sense that quantum phenomena exhibit correlations between  
  spacelike separate measurements that are inconsistent with
  any common cause explanation.  Nevertheless, many authors 
  state that
  this odd nonlocality could not involve any controllable
  superluminal transmission of momentum-energy, signals, or information, 
  since there are several proofs in the literature apparently showing 
  that the expectation value of any observable at one location in a 
  phase-entangled multi-particle system cannot be affected by any choice 
  of measurement strategy employed on some other spacelike-separate part 
  of the system.  However, we claim that
  most or all no-signalling proofs published to date are
  question-begging, in that they
  depend upon assumptions about the locality of the dynamics
  of the measurement
  process that are the very points that need to be established in the
  first place.  In this paper, we undertake 
  a critical examination of
  no-signalling proofs by Bohm and Hiley \cite{BH93} 
  and Shimony \cite{Shimony1}, which
  illustrate the problem in an especially striking way.
  }
}

\newpage
{\baselineskip=14pt
\begin{quote}
  Pronouncements of experts to the effect that something cannot be done
  have always irritated me.
  \rightline{---Leo Szilard \cite[p. 28]{Rhodes}}
\end{quote}
\medskip
\begin{quote}
  It seems to me that it is among the most sure-footed of quantum 
  physicists, those who have it {\em in their bones}, that one finds the 
  greatest impatience with the idea that the `foundations of quantum 
  mechanics' might need some attention.  Knowing what is right by 
  instinct, they can become\nolinebreak\hspace{2pt}
  a little impatient with nitpicking
  distinctions between theorems and assumptions.
  \rightline{---J. S. Bell \cite[p. 33]{Bell1}}
\end{quote}

\section{INTRODUCTION}
Let us imagine a typical EPR (Einstein-Podolsky-Rosen) experimental 
scenario, in which a centrally-located source is sending out pair after 
pair of correlated particles,
which we shall label $A$ and $B$,
in opposite directions. \cite{EPR,Bohm1}
At equal distances
from the source we shall suppose that there are two detectors, $D_A$ and 
$D_B$, at rest with respect to the source.  (We make these stipulations
to evade the considerable complications entailed by relative motions of 
source and detectors.)  At $D_A$ and $D_B$ sit Agents Mulder and Scully
respectively, patiently writing down the results of each 
run of the apparatus.
Mulder is holding his detector at a constant angle, while
Scully varies her detector angle from time to time, hoping to send a 
message to her partner.

We know that the results recorded by Mulder and Scully will be 
correlated. To be a bit more specific, if the particles are fermions
of spin $1/2$,
and if we are recording spin up or down in a particular direction, then
the correlation will be given by $-\cos\theta_{AB}$, where $\theta_{AB}$
is the relative angle between the two detectors.  
We know that this
correlation violates a Bell Inequality
\cite[p. 140--147]{BH93},
and we know that this means
that the particular results our two agents get could not have been 
encoded in the particles when they left the source. \cite{Itamar}
But we also
know that Scully's attempts to communicate with Mulder directly will be 
thwarted, for no matter what manipulations she performs on her
detector, all that either she or Mulder will record will be an 
apparently {\sl random} sequence of ups and downs.  Only when the two 
sequences of results
are compared at a later time, will it be seen that 
correlations stand between them, satisfying the above formula.

The best that Scully can do is
impose a signal upon the {\sl correlations} by
varying her detector angle; and indeed, this would make possible,
in principle at least, the most theoretically perfect encryption
scheme that one
could imagine.  Either agent's string of random results would serve as 
the unique key for the other, and eavesdropping could be detected by a
tendency of the results to {\sl obey} a Bell Inequality (since 
eavesdropping destroys the correlations).
\cite{Lomonaco}
But there does not seem to be any way 
that Scully can send a message that shows up in Mulder's local 
statistics.  If all she does is adjust her detector angle, Mulder just 
continues to see what looks like random noise.  If, on the other hand, 
Scully interposes some magnets or other devices to force the
particles to go
through her detector in a particular direction, she will 
discover later on, after the results are compared, that not only does 
Mulder continue to receive random noise, but that she has also
washed out the correlations.

The relativistic prohibition against superluminal signalling thus seems 
to be protected.  However,
Mulder is still puzzled, because he is swayed by
Tim Maudlin's very
persuasive arguments that the violation of the Bell Inequalities in 
experiments such as this can only be accounted for by the assumption 
that there is some sort of superluminal causation, in apparent violation 
of the theory of relativity.  \cite{Tim}  
Mulder is well aware that if there were
something Scully could do that would preserve the
correlation between their results,
but at the same time allow her to control which way her
particles go,  then this
would not only threaten causal
paradoxes, but would allow him and Scully to synchronize their watches
instantaneously and thereby violate Einstein's relativity of
simultaneity.       But he quite fails to see why this might not, in 
principle at least, be possible.
Finally, in utter frustration, Mulder
concludes that there is a hidden conspiracy
between quantum mechanics and relativity, such that relativity will 
always appear to be obeyed even when it is being covertly violated.


The ever-sensible Scully will assure Mulder that 
things are just as they should be,
since numerous authors have
published proofs demonstrating,
or supposedly demonstrating,
that no-controllable-signalling is a
completely general property of quantum mechanics.  \cite{Eberhard, 
ER, Shimony1,GRW,BJ,Jordan}
However, Mulder, never content merely to
accept the authority of experts, reads some of this literature, and begins
to develop suspicions about the logical pedigree of the widely-cited
proofs it contains.  In this paper,
we will put two especially pertinent
examples of no-signalling
proofs under the microscope, and show that Mulder's worries are
justified.

\section{Can We Explain the Correlations?}
To place the discussion of signalling in context, we will
consider two strongly contrasting approaches
to the following  question:
How can we explain correlations between spacelike-separate
events, when recourse to a common cause is ruled out?  

\begin{enumerate}
\item{``Don't ask''.  One notes that we already have an empirically
  adequate set of algorithms
  for predicting observable correlations, and
  combines this fact
  with the warning of Bohr \cite{Bohr35} that to ask for a
  spatio-temporal account of the interactions between correlated 
  particles is to ask an experimentally ill-posed question.  As David 
  Mermin puts it,
  \begin{quote}
  My own view on EPR which keeps changing---I offer this month's 
  version---is that barring some unexpected and entirely revolutionary 
  new developments, it is indeed a foolish question to demand an 
  explanation for the correlations beyond that offered by the quantum 
  theory.  This explanation states that they are the way they are 
  because that's what the calculation gives. \cite[p.202]{Mermin}
  \end{quote}
  This very Humean view has it that there is no basis for belief in
  ``hidden powers'' or
  ``necessary connexions'' between events.
  The price we have to pay for
  the huge predictive effectiveness of quantum mechanics, is, in effect,
  to give up
  the hope of understanding the actual basis of physical phenomena.
  }
\item{One accepts that if there is any sensible explanation of the 
  correlations at all, then it must involve
  some sort of direct (and therefore
  superluminal) causal interaction between the distant particles.
  Tim Maudlin puts it bluntly:
  \begin{quote}
  Bell concluded that violations of the inequality demonstrate that the 
  world is not locally causal, i.e., that these phenomena cannot be 
  reproduced by any theory which postulates only locally defined 
  physical states which cannot influence states at space-like 
  separation\dots  Philosophers of physics have been wont to question 
  this conclusion\dots Bell was, however, quite correct in his 
  analysis.  Statistics such as those displayed by the photons [in an 
  EPR scenario] cannot be reliably reproduced by any system in which the 
  response of each particle is unaffected by the nature of the 
  measurement carried out on its distant twin.  The photons remain ``in 
  communication'' no matter how great the spatial separation between 
  them.  Instead of trying to deny these non-local (i.e., superluminal) 
  influences, we should begin to study the role such influences must 
  play in generating the phenomena. \cite[p. 405]{Tim2}
  \end{quote}
  }
  We take the notion of studying ``the role such influences must play in
  generating the phenomena'', to mean that we should find out
  what features of a
  theory of superluminal influences would be {\em necessary} in order to
  reproduce the
  observed behavior.  As we shall see below, there is one class of 
  candidate theories---the causal interpretations of quantum mechanics 
  proposed by Louis de Broglie and David Bohm---that are apparently {\em
  sufficient} to account for the observed phenomena.  However, we still
  do not know how much choice we have in adopting such theories.
\end{enumerate}

These are only two of the many attempted interpretations of QM,
some of which are of
great subtlety and ingenuity.  However, it is not too much of an 
exaggeration to say that most interpretations of QM
are aimed at finding some way of
{\sl accepting} the nonlocality implied by Bell's Theorem---which, as noted
above, is essentially a negative result, amounting to the elimination of 
common-cause explanations of quantum correlations---{\sl without} going 
as far as alternative 2 contemplates; that is, without 
swallowing the idea that
one particle {\sl literally} exerts an instantaneous influence on
its distant partner.  Hence, it is useful to focus on these two views, 
since they represent
two extremes of thought on the problem.

Note carefully that a supporter of position 1 (above) could say that
there is a {\it non sequitur} in Maudlin's argument: from the fact
that no local explanation is available, it does not follow that
some other sort of explanation is possible.  
It might well be that there is no
explanation at all; in other words, that the Bell-Inequality-violating
correlations of QM are simply basic, 
raw data that are the {\sl starting points} for any full 
development of
physics, not something that could be explained by any deeper physical 
theory.  (This has been proposed, for instance, by Fine \cite{Fine} and
Pitowsky \cite{Itamar}.)
A defender of position 2, therefore, will ideally have to show 
that there are other motivations for considering non-local causation, 
apart from the fact that it would furnish a {\it prima facie}
explanation for the
correlations.  And, indeed, supporters of the Bohm/de
Broglie alternatives do have some grounds to claim that their theories 
are broadly motivated by the mathematical structure of wave mechanics.

The ``don't ask'' option is widely endorsed, especially by many working
physicists. It does have the advantage that it
tends to keep one out of trouble, and this has some survival value
in today's scientific
ethos, according to which it is impermissible to be perceived to have
made a mistake.\footnote{At the risk of over-stating the obvious, we 
  believe that this aspect of the contemporary scientific ethos is
  counter-productive.}
Furthermore,
option 2 has been long
regarded by many as outside serious discussion
both because it leads to possible
conflicts with relativity, and because of a deeply-felt instinct that 
physics should be local.
Einstein himself
dismissed the notion of nonlocal causation as ``spooky action 
at a distance''.

An important difference between answers 1 and 2,
is that according to the latter, there is {\sl new physics} to be
uncovered; while according to 1 there is no reason to suppose that the 
present formulation is not as good a theory as we are going to get.  
According to 1,
nonlocality would not be
something one understands, but something to which one
adjusts.  Interpreting QM would be a typical case of
what Wittgenstein famously called ``letting the fly out of the 
fly-bottle''---seeing that if only we think about a problem
the right way,
there is no problem at all.  It must be said that this position, while 
logically open given our present state of knowledge, is most
uninteresting, since it virtually guarantees that our understanding will 
not move much beyond its present state.

\section{Causal Interpretations  of QM}
Despite long-standing prejudices against taking the idea of superluminal 
or nonlocal causation seriously,
there is increasing recognition that
the causal interpretations inspired
by the theories of David Bohm \cite{Bohm2} and
Louis de Broglie \cite{Louis1,Louis2} are
among the best contenders to provide a deeper explanation, if not a
generalization, of QM.
The central feature of such theories is that they countenance some sort
of direct dynamic interaction between
correlated particles.\footnote{
   There is a recent variant of Bohm's theory known as
   ``Bohmian Mechanics,'' in which particle motions are 
   supposed to be correlated by a
   sort of pre-established harmony. We will not consider that here,
   save to note that it is subject to the same objections
   to any theory with a local Hamiltonian,
   that we raise in the next section.  
   For
   a superbly perspicuous overview of the various flavours of the causal 
   interpretation, see \cite{Cushing}.
   }
Bohm's theory (which is much more widely studied) 
can be considered to be a non-relativistic approximation to the 
relativistic theory of de Broglie.  In Bohm's theory,
interactions between particles 
are mediated by a mysterious potential having the form
\begin{equation}
  Q = { \frac{\hbar^2}{2m} } {\frac{\nabla^2 R}{R} }
\end{equation}
where $m$ is the particle mass, and $R$ is the amplitude of the wave 
function
\begin{equation}
  \Psi = R \exp(iS/\hbar).
\end{equation}
(The quantity $S$ is the action of the system.)  
In the case of phase-entangled multiparticle systems, the quantum
potential for the system cannot, in general, be written merely as the 
sum of the quantum potentials for the individual particles.  Rather, it 
is a global property of the system as a whole.  (See \cite[p. 
62--63]{Cushing}.)
The quantum potential
contributes to
the total mass-energy of a multi-particle system, and, when 
differentiated with respect to distance, defines a force---literally, a 
sort of action at a distance---that Bohm frequently argued would
be a natural way to account
for the correlations between distant particles.

There are many questions to be asked about the best way to interpret and 
develop the insights of Bohr and de Broglie.  The crucial point to 
grasp, though, is that the quantum potential
$Q$ is by no means an arbitrary construct, but something that 
can be derived
straightforwardly from certain basic assumptions of wave mechanics.  
(See \cite{Bohm2,Cushing,Louis1}, or many other sources.)  Option 2 is,
therefore, to be taken very seriously, both because (as Maudlin insists) 
it seems, {\it prima facie} at least,  to be demanded by the
observed
failure of the Bell Inequalities, and also because
something like the theories of de Broglie or Bohm
have been implicit in
the mathematical structure of quantum theory from the outset.
But this makes the question of signalling especially
acute, as we shall see.

\section{Bohm and Hiley on Signalling}
In their {\sl Undivided Universe}  \cite[Chapter 7]{BH93},
David Bohm and Basil Hiley attempt to address the
problem of superluminal signalling in quantum mechanics.
Our claim will be that their argument is question-begging, since,
as we shall see,
they rule
out of consideration from the beginning the very possibility they most 
need to examine --- {\sl especially} given their stated commitment to causal
interpretations of QM.

The charge of circularity has already been leveled
against a large class of no-signalling proofs within non-relativistic 
quantum mechanics and local quantum field theory
by J. B. Kennedy \cite{JBK}, and also by one of us
\cite{KP1,KP2}.  The value in studying this particular
argument by Bohm and
Hiley is that they express in a remarkably clear
form the fallacy that is typical of virtually all the no-signalling 
arguments with which we are familiar.  We say this in all 
due respect for these authors, who have made great contributions to
physical science.  (It is, in particular,
a disgrace that Bohm, like J. S. Bell, was not awarded
  the Nobel Prize in Physics.)
Our claim is not that they have been
especially careless,
but that, given the long-standing commitment of science to
locality, theirs is a remarkably easy mistake to make.

In discussing various possible interpretations of
the EPR experiment, they
remark,
\begin{quote}
  \dots it seems very reasonable to suggest that $A$ and $B$ [the 
  spacelike separate particles] are directly connected, though in a way 
  that is perhaps not yet known. \cite[p. 139]{BH93}
\end{quote}
This is essentially a variant of alternative 2, above, and it is,
indeed, the central claim of causal accounts of QM such as the
theories of Bohm and de Broglie.  The ultimate problem, of course, is to
elucidate the nature of the ``connection'' between the particles.

However, they then set out to immediately scotch any fears that such
hypothetical direct connections,
whatever they might look like in detail,
could be
used to signal superluminally.  
Their argument is given in wave-mechanical terms;
what follows here is their derivation re-expressed in the more
perspicuous Dirac notation.

We shall suppose
that ``an external system [measurement device] with coordinate $y$ is 
allowed to interact with the spin of particle $A$.''  
The initial state vector for a system of
two spin-coupled particles $A$ and $B$,
and a measuring apparatus with coordinate $y$, will be
\begin{equation}
  \ket{\psi_{0}} = \ket{\phi_{0}^{y}} \squarthtwo [\ket{+_{\a}^{A}}
  \ket{-_{\b}^{B}} - \ket{-_{\a}^{A}}\ket{+_{\b}^{B}} ]
\end{equation}
where the superscripts $A,B,y$ indicate the Hilbert spaces for particle
$A$, particle $B$, and the measuring apparatus, respectively.  
The subscripts $\a$
and $\b$ indicate the spin direction for which the 
${\ket{+},\ket{-}}$ is a basis set,
and the ket $\ket{\phi_{0}^{y}} $ represents the initial wave function for
the measuring device. (The ket products are to be understood as direct 
products, although we have dropped the usual   $\otimes$ notation).


An interaction between the measuring device and the spin of particle $A$ 
is then ``carried out''.   The immediate question is how we should 
represent this.

Here is the key passage:
\begin{quote}
  The most general possible result of this interaction will be 
  represented by a unitary transformation on the subsystem consisting of 
  $y$ and $A$, because, {\sl by hypothesis}, [our emphasis], we are
  assuming our interaction does not directly disturb $B$.  If it did 
  then this would not constitute sending a signal from $A$ to $B$, but 
  would just be a direct disturbance of $B$ by its interaction with 
  $y$.\cite[p. 139]{BH93}
\end{quote}
Bohm and Hiley then go on to show
that given this assumption there is no change in
the expectation value of the spin operator for particle $B$ as a
consequence of the measurement made on $A$.  
We will comment, below, on the cogency of the reasoning expressed in
this passage.  First, though, we summarize the calculation.

We represent such a unitary transformation by the operator
$U_{\a,\a^\pr}^{A,y}$ where the superscripts indicate that this operator only
works on the Hilbert spaces of the apparatus and particle $A$ and 
the subscripts show that it performs
the operation of rotating the initial basis states of $A$
from the direction $\a$ to 
$\a^\pr$.  The state of the system then becomes
\begin{equation}
U_{\a,\a^\pr}^{A,y}\ket{\psi_{0}} =\squarthtwo
\ket{U_{\a,\a^\pr}^{A,y}\phi_{0}^{y}}
[\ket{+_{\a^\pr}^{A}}\ket{-_{\b}^{B}} -
\ket{-_{\a^\pr}^{A}}\ket{+_{\b}^{B}} ].
\end{equation}

By assumption, the basis kets of $B$ are unaffected by this transformation.
Bohm and Hiley
then go on to show, unsurprisingly, that given this assumption there is
no change in 
$\expect{\sigma_{\b}}^\pr$,
the new expectation value of the spin operator 
(in direction $\beta$)
for particle $B$ 
as a consequence of the measurement made on $A$.  We write
\begin{eqnarray}
\expect{\sigma_{\b}^{B}}^\pr
  & =  &
  \matrix{ U_{\a,\a^\pr}^{A,y}\psi_{0}} {\sigma_{\b}^{B}}
  {U_{\a,\a^\pr}^{A,y}\psi_{0}}  \nonumber  \\
& = & 
  \half{\braket{U_{\a,\a^\pr}^{A,y}\phi_{0}^{y}}
  {U_{\a,\a^\pr}^{A,y}\phi_{0}^{y}}}  \nonumber \\
&   &
  [\bra{+_{\a^\pr}^{A}}\bra{-_{\b}^{B}} -
                \bra{-_{\a^\pr}^{A}}\bra{+_{\b}^{B}} ]
 \sigma_{\b}^{B}
 [  \ket{+_{\a^\pr}^{A}}\ket{-_{\b}^{B}} -
                \ket{-_{\a^\pr}^{A}}\ket{+_{\b}^{B}}  ]
\end{eqnarray}
Since the orthonormality of the states is retained under a unitary
transformation, and since $\sigma_{\b}^{B}$ operates on particle $B$ alone
(as if it ``passes through'' the $A$-kets),
this gives
\begin{equation}
\half[\matrix{-_{\b}^{B}}{\sigma_{\b}^{B}}{-_{\b}^{B}} -
\matrix{+_{\b}^{B}}{\sigma_{\b}^{B}}{+_{\b}^{B}}]  =
\expect{\sigma_{\b}^{B}}.
\end{equation}
To sum up:  since {\sl ex hypothesi} the unitary transformation only
operates on the Hilbert spaces of the measuring device and particle $A$, the
expectation value for the spin of particle $B$ is the same before and after the
interaction.

Several comments come to mind.
First, this whole line of reasoning is very odd,
since the authors only a few lines above on the same page readily 
concede that
$A$ and $B$ may be ``directly connected'', and it is hard to see how, if 
this were so, something done to $A$ might not produce a ``direct 
disturbance'' of $B$.  (Presumably, ``direct'' means ``nonlocal'', at
least in the sense of being instantaneous, or not
involving only retarded
reactions.)  Bohm and Hiley therefore seem to contradict 
themselves; they insist
on the plausibility of a direct connection between the particles, but 
then describe the situation in a way that excludes that very possibility.

Does their proof amount to anything more than an illustration of the 
fact that an operator that doesn't operate on a wave-function doesn't
change the wave-function?  (Kennedy argues that virtually all 
no-signalling arguments within nonrelativistic quantum mechanics boil 
down to this unexceptionable claim,
at least mathematically.  \cite{JBK})
That would not seem to be especially
illuminating.

Here is a more charitable reading:  even though proofs of this sort 
cannot show that there is no direct causal interaction between left and
right particles, they do show that there is
no inconsistency in the formalism of quantum mechanics, such
that we would get evidence of a superluminal causal interaction if we 
{\em assume} there is none.  In other words, one cannot
beat the house merely by some sort of statistical trickery.

It was, no doubt, a salutary exercise
to have shown this, but
the use of such a calculation in support of a general no-signalling 
claim is completely question-begging.
This is because
it is very hard to see how any sort of
signal from $A$ to $B$ would not require the disturbance of $B$ by $A$, 
albeit in some fashion ``that is perhaps not yet known''.

This point requires special emphasis.  It is a basic result of
information theory that any form of information transmission requires 
the expenditure of free energy.  The reason is that to encode 
information in a physical
structure (for instance, to do something that causes
a measurement device to display some definite outcome) is to lower the
entropy of that structure.  There are many
ways in which this can be accomplished, but all
require the doing of some work on that structure.
Transmission of information from $A$ to $B$ without direct 
disturbance---{\sl whether controllable or not}---would
be a violation of the Second Law of Thermodynamics, since one would have
achieved an energetically free reduction in entropy.
Therefore, to
suppose that one could signal without ``direct interaction'' is to
misunderstand the nature of signalling in general.

In other words, the most that the no-signalling argument by Bohm and Hiley
really shows---and this is true of all the no-signalling arguments we 
cite above, and most in the literature\footnote{
  A. Valentini has a highly original treatment of the signalling problem 
  in his own version of Bohmian Mechanics.  \cite{Valentini1}  
  Valentini, following Bohm and Vigier \cite{BV},
  treats the equation $P(x) = |\Psi(x)|^2$, which he dubs the ``quantum 
  equilibrium'' condition,
  not as a mathematical
  identity, as it is in the standard abstract formulation of quantum 
  mechanics, but as a thermodynamic average which could have been 
  violated in the early universe.  Valentini shows that, in his
  theory, no-signalling
  holds so long as quantum equilibrium holds.  Whether or not 
  Valentini's approach is sound, it is less obviously question-begging 
  than the usual no-signalling arguments.  However, all 
  presently extant versions of
  Bohmian Mechanics assume a local Hamiltonian for the multi-particle
  system, and are thus open to objections we raise in the next section.
  }
---is that the quantum
mechanical measurement process cannot be used to violate the Second Law 
of Thermodynamics.  One cannot signal by sheer sympathetic
magic; that is, {\em without}
actually, physically interacting with the receiver.  
However, these arguments utterly fail to show whether or not there 
{\em exists} a direct
interaction between the distant particles, even though
this is precisely the point that is at issue.
It is not relativity
that is protected by the no-signalling
arguments, but thermodynamics.

\section{Nonlocality of Multiparticle Dynamics}
It will be instructive to take a closer look at the widely-cited
no-signalling argument by Abner Shimony \cite{Shimony1}, which
(by using the Hamiltonian formalism) explicitly considers the dynamics 
of ``entangled'' states.

Shimony invites us to consider an EPR scenario with correlated particles 
$A$ and $B$.  
We want to write the Hamiltonian for this
system, in the case that a measurement 
device $D_B$ acts on $B$.
Shimony assumes that this total Hamiltonian can be
written in the form
\begin{equation}
  H_{\rm tot} = H_A \otimes {\bf 1}_B + H_{DB} \otimes {\bf 1}_A, 
  \label{Shimony}
\end{equation}
where $H_A$ is the Hamiltonian of particle $A$, ${\bf 1}_A$ is the
identity operator on $\Hilbert{A}$,
the Hilbert space for $A$ (and similarly for
${\bf 1}_B$), and $H_{DB}$ is the Hamiltonian of 
the combined system of
$D_B$ and particle $B$. Adopting a Hamiltonion of this form
amounts to assuming dynamic locality at {\sl two} levels:
\begin{description}
\item[S1]{It
  assumes that $D_B$ interacts only with $B$;}
\item[S2]{It assumes that the
  combined system of $D_B$ and $B$ does not interact with $A$.}
\end{description}

These 
assumptions do seem to be perfectly reasonable given normal classical 
intuitions about how particles interact, since we would assume that once 
the particles are sufficiently far apart, any immediate reactions 
between them would drop rapidly to zero.  (There could be retarded
interactions, of course, but here we are only concerned with what 
happens at some definite time in the lab frame of reference.)  However, 
in the context of this investigation, we
are not entitled to rely upon such classical intuitions, because the 
entire point is to see whether or not they are sound.

In any case, given Eq.~\ref{Shimony},
one can show (by series expansion) that the time evolution operator for
the total system factorizes:
\begin{eqnarray}
  U(t) & = & e^{iH_{\rm tot}t} \\
       & = & e^{iH_At} \otimes e^{iH_{DB}t}.
\end{eqnarray}
Shimony then sets out to calculate the expectation value of some 
operator $G$ acting on particle $A$ alone, given this action of $D_B$ on 
$B$.   If any such measurement carried out  on  $B$ can influence the 
expectation value of any observable measurable on $A$, then Scully can,
indeed, signal to Mulder, by varying the parameters of the apparatus 
$D_B$.

We first need an expression for the total system state.  Let $\ket{a_i}$ 
be basis states for $\Hilbert{A}$, and $\ket{b_i}$ be basis states for
the Hilbert space $\Hilbert{B}$ of particle $B$.
The assumption that
$D_B$ acts dynamically on $B$ alone implies that we can represent the 
effect of $D_B$ on the total system in terms of operators acting 
strictly on a Hilbert space 
$\Hilbert{B}^\pr = \Hilbert{D_B} \otimes \Hilbert{B}$,
where $\Hilbert{D_B}$ is the Hilbert space of the measurement
apparatus.  Writing the basis states of
$\Hilbert{B}^\pr$ as $\ket{b^{\pr}_i}$, the state of the total system
(apparatus plus entangled particles $A$ and $B$), at time $t_0$,
can be written as
\begin{equation}
  \ket{\psi(t_0)}  =  \sum c_i\ket{b^\pr_i a_i}.
\end{equation}
Clearly this is not, in general,
factorizable---even though we are assuming that its
time evolution is!

After a time $t$ the system has evolved to a state
\begin{equation}
  \ket{\psi(t)}  =  U(t - t_0) \ket{\psi(t_0)}.
\end{equation}
As with Bohm and Hiley's calculation, we are assuming that the 
measurement interaction with $D_B$ does not collapse (i.e., project) the 
state, but evolves it in a unitary way.

To calculate $\expect{G}$, we observe that $G$'s action on the global 
system can be represented by $G_{\rm tot} = G \otimes \Id_B$.  Then we get
\begin{eqnarray}
   \expect{G_{\rm tot}} & =  & \matrix{\Psi(t)}{G \otimes 
   {\Id_B}}{\Psi(t)} \\
   &  =  &  \matrix{\Psi(t_0)}{U^\dag(t-t_0) (G \otimes {\Id_B})
   U(t-t_0)}{\Psi(t_0)} \\
   &  =  &  \matrix{\Psi(t_0)}
   {(e^{-iH_A(t-t_0)} \otimes
   e^{-iH_{DB}(t-t_0)})
   (G \otimes {\Id_B}) \nonumber \\
   &     &
   (e^{iH_A(t-t_0)} \otimes
   e^{iH_{DB}(t-t_0)})}
   {\Psi(t_0)} \\
   &  =  &  \matrix{\Psi(t_0)}
     {(e^{-iH_A(t-t_0)} G e^{iH_A(t-t_0)})
     (e^{-iH_{DB}(t-t_0)} {\Id_B} e^{iH_{DB}(t-t_0)})}
       {\Psi(t_0)}  \\
   &  =  &  \matrix{\Psi(t)}{G}{\Psi(t)}  \\
   &  =  &  \expect{G}
\end{eqnarray}
In the end, $\expect{G}$ shows no dependency on whatever may have been 
done on particle $B$.
In other words, since $A$ and $B$ are presumed causally independent, a
measurement on $B$ cannot influence the statistics of measurements on 
$A$.  This is, of course, just a more general version of the argument of 
Bohm and Hiley.

Abner Shimony himself is well aware
of the relevance of the dynamics for the
signalling problem.  Elsewhere, he states,
\begin{quote}
  \dots quantum mechanical predictions concerning ensembles of pairs of 
  particles do not violate Parameter Independence [no-signalling], 
  provided that nonlocality is not explicitly built into the interaction 
  Hamiltonian of the particle pair.  \cite[p. 191]{Shimony2}
\end{quote}
Evidently, Shimony did not believe that there was any physical
justification for 
considering explicitly nonlocal Hamiltonians.  However, we need only 
look a few pages ahead in Bohm and Hiley's book to see that there is.

\section{Symmetrization and Nonlocality}
In a section of {\sl The Undivided Universe} entitled ``Symmetry and 
antisymmetry as an EPR correlation,'' \cite[pp. 153--157]{BH93}
Bohm and Hiley point out that wave functions 
of multi-particle systems
may be symmetric or
antisymmetric.  Particles belonging to systems
with symmetric wave functions
exist in identical states, and accordingly
obey 
Bose-Einstein statistics,
while particles with antisymmetric wave
functions obey Fermi-Dirac statistics, and must obey  the exclusion 
principle.

Suppose our particles $A$ and $B$ are bosons.  We wish to measure some 
operator $O_A$ on particle $A$.  There will have to
be a corresponding operator 
$O_B$ acting on $B$, since $A$ and $B$ must obey identical statistics.
Therefore, as Bohm and Hiley explain
(\cite[p. 153--154]{BH93}),
in order to
maintain the symmetry of the Hamiltonian between the two particles, we 
must write the Hamiltonian of the measurement interaction as
\begin{equation}
  H^S_I  =  \lambda(O_A + O_B)\dee{}{y}   \label{nonlocal}
\end{equation}
This obviously violates assumption S1 above, 
because of the dependence upon $O_B$,
and thus renders
Eq.~\ref{Shimony} entirely inapplicable.  It also, again obviously,
contradicts the behavior of the unitary
transformation used by Bohm and Hiley only a few pages earlier in their 
own book.  We note, also (a point not explicitly mentioned by Bohm and 
Hiley), that, as far as we know, {\em all} particles are either bosons 
or fermions, and must therefore obey symmetrization conditions.  The 
best we can say, therefore, is that the whole treatment of signalling 
typified by the Bohm-Hiley and Shimony proofs could only be applicable 
in cases in which these symmetrization conditions can be ignored.

Nothing we have said here shows that systems with nonlocal
Hamiltonians such as Eq.~\ref{nonlocal}
could, indeed, be used for controllable signalling.
However, proofs of the type offered by Shimony, or Bohm and Hiley,
are clearly powerless to show that they cannot.

Finally, observe that one has to 
use the nonlocal Hamiltonian of Eq.~\ref{nonlocal}
{\sl whether or not} one
accepts a causal interpretation of QM.  As Bohm and Hiley carefully 
note, we have to use a symmetrized
Hamiltonian like this if we want to get the right
predictions for Bose particles, and that fact is
quite independent of whatever
interpretation of QM one chooses.  Hence, the question of signalling is, 
in the last analysis, just as unavoidable for Option 1 as for Option 2.

The notion of nonlocal energy is, admittedly, difficult to grasp.  One 
might be inclined to think that according to a causal interpretation,
there must be some sort of
superluminal transmission of a localized pulse of mass-energy between 
the remote particles.  However,
if we ask whether energy is being shuttled superluminally between $A$
and $B$, by tachyons perhaps, we miss the point.  Some such description 
might be useful in some contexts.  However, the real point is that 
mass-energy is {\em nonlocal}, a {\em global} property of a 
multi-particle system.  The multiparticle
system as a whole will have a spectrum of 
possible energy states, and the energy is not any place in particular at 
all; it is just a general property of the system, that may make itself 
manifest in a variety of ways.  
It is probably safe to say that this is analogous to the way in which 
the energy of an electron orbital in an atom is a global property of the 
orbital as a whole.
Localization of mass-energy is a
process that happens in certain specific circumstances that we do not 
fully understand as yet.

Our remarks here can only serve to indicate some enticing possibilities.
The important point to note is that there are numerous indications 
within quantum physics that the dynamics of multiparticle systems are, 
in general, nonlocal.  It seems to be largely philosophical prejudice 
against nonlocality that has prevented us, so far, from following up on 
these leads---a philosophical prejudice against which Bohm and Hiley 
argue persuasively (\cite{BH93}), but 
to which they appear to have fallen victim themselves.

Any satisfactory treatment of the signalling problem must employ a 
formalism that {\em explicitly} takes into account the possibility of 
nonlocal causal interactions.  Exactly how we should do this remains to 
be seen, although the causal versions of quantum mechanics
of de Broglie and Bohm offer promising leads.  Bohm's
interpretation,
however, suffers from the possible
defect that it takes as a starting point the 
Hamiltonian
\begin{equation}
  H = -\frac{\hbar^2}{2m}\nabla^2 + V.
\end{equation}
(See \cite[p. 28]{BH93}.)
The first term represents the kinetic energy of the particle, and the 
second represents local potentials such as electromagnetic potentials.  
It is likely that this Hamiltonian represents some sort of 
semi-classical limiting approximation, not the accurate potential for a 
system of correlated particles.  However, this point requires much 
further investigation.\footnote{
  Deotto and Ghirardi (\cite[p. 24]{DG} briefly consider a Hamiltonian 
  with the quantum potential written in.
  }

In the end, we can safely say to Agent Mulder that there is no hidden 
conspiracy, but merely a confusion.  Except for systems in which the 
Hamiltonian approximates to a local form, as in Eq.~\ref{Shimony},
we still
simply do not know whether one
can violate relativity by means of some sort of controllable nonlocal 
effect in entangled multiparticle states.  The truth is still out there.

\section*{Acknowledgements}

This work was supported by the Social Sciences and
Humanities Research Council of Canada and the University of Lethbridge.  
K. P. thanks James Robert Brown for valuable discussions and guidance
in the early stages of this research.

%
%


{
}   

}   

\end{document}